\documentclass[aps,prb,twocolumn,groupedaddress]{revtex4}
\usepackage{graphicx}
\begin{document}

\title{EPR study of superconductors}



\author{I. A. Garifullin}

\affiliation{Zavoisky
Physical-Technical Institute, Kazan Scientific Center of Russian
Academy of Sciences, 420029 Kazan, Russia}


\date{\today}

\begin{abstract}

Historical review on the studies of the electron paramagnetic
resonance in superconductors performed in the period from 1970 to
1990 at the Kazan Physical Technical Institute of the Russian
Academy of Sciences (group of Dr. E. G. Kharakhash'yan) in
collaboration with Kazan State University (group of Prof. B. I.
Kochelaev) and with the Institute for Physical Problems of Russian
Academy of Sciences (group of Prof. N. E. Alekseevskii) is
presented. We have observed for first time electron paramagnetic
resonance of impurities in a type II superconductor; found
indication for a long-range exchange interaction between magnetic
impurities arising due to the superconducting correlations; observed
the magnetic ordering of impurities in the superconducting state;
and, finally, we found one of the first evidences for heterogeneity
of the 1:2:3 high-$T_c$ superconductor which is its natural property



\end{abstract}
\maketitle

\section{Introduction}
\label{intro} Superconductivity is a striking physical phenomenon;
it is a manifestation of quantum effects on a macroscopic scale. The
most striking features of superconductivity are (i) the zero value of
the electrical resistivity, (ii) the expulsion of the magnetic flux
(Meissner effect), (iii) the interference between macroscopically
separated junctions. None of these phenomena can be studied using
any magnetic resonance method because the basic properties of
superconductivity are the quantum effects on a macroscopic scale,
while magnetic resonance is a microscopic tool. Therefore, the
properties of superconductors from magnetic resonance were not
expected to be striking as the above mentioned effects. However, it
was expected that they can help to investigate the microscopic
aspects of superconductivity. In particular, these are the
superconducting coherence phenomena, the superconducting and
magnetic correlations.

Significant role of the nuclear magnetic resonance (NMR) studies of
superconductors is well known. Indeed, one of the first experimental
confirmations of the validity of microscopic  theory derived by
Bardeen-Cooper-Schriffer (BCS) \cite{BCS} was obtained by NMR. Hebel
and Slichter \cite{Slichter} have shown experimentally that the
nuclear spin relaxation time $T_1$ just below the superconducting
transition temperature $T_c$ is shorter than in the normal state.
That is the direct evidence of validity of BCS theory. This happens
(see, e.g., \cite{Weger}) due to the opening up of the
superconducting gap when the total number of states is a constant,
the density of states of conduction electrons at the Fermi level
$\rho ({E_F})$ averaged over an energy interval of about $k_BT$ (for
$T\simeq T_c$) is larger in the superconducting state than in the
normal state. Therefore, the shortening of $T_1$ just below $T_c$ is
not surprising; the salient feature of the BCS theory is that for
spin independent phenomena such as ultrasonic attenuation, this
effect does not occur since the matrix element for the transition
becomes smaller and cancels the density of states effect, while for
relaxation due to the contact interaction the effect is present.
Strictly speaking, according to the BCS theory
$1/T_1\rightarrow\infty$ at $T=T_c$ since density of states $\rho
({E_F})$ is finite, but $\int [\rho (E)]^2 dE$ diverges; since the
BCS theory is only an approximation, and the excitation in reality
have a finite lifetime, $1/T_1$ does not become infinite. Hebel and
Slichter \cite{Slichter} incorporated this lifetime effect by
empirical smearing of the density of states function.

It is necessary to note that in the first works on NMR in
superconductors long nuclear relaxation time allowed to overcome the
main difficulty for the observation of the resonance caused by the
Meissner effect using the method of cycling of dc magnetic field
\cite{Slichter,Redfield}. Formation of nonequilibrium response of
the nuclear spin system was performed in the superconducting state
and the resonance signal was observed in the normal state. Because
of a short relaxation time this method can not be used in the case
of EPR. For a long time the possibility of using EPR of conduction
electrons for the study of the superconducting state was considered
to be impossible since the Cooper pairs which are formed by two
electrons with opposite spins in the superconducting state have the
zero total spin. Furthermore, superconductors expel completely the
external magnetic field from their inside due to the Meissner
effect. This makes impossible even the observation of the resonance
of a localized moment which can be specially introduced into the
superconducting matrices as a special spin probe. However, in type
II superconductors in the mixed state magnetic field penetrates as
the Abrikosov vortices \cite{Abrikosov}. This offered a possibility
to use EPR for the study of the type II superconductors.

The works concerning the study of superconductors by EPR were
started more than 40 years ago when its pioneering observation by
the group of E. G. Kharakhsh'yan \cite{Altshuler} was done at the
Kazan Physical Technical Institute of the Russian Academy of Sciences
(in present Zavoisky Physical Technical Institute). Soon after that
two groups possessing the most sensitive in the world EPR equipment
which allow to perform measurements at low and ultralow
temperatures, join in the study of EPR in superconductors. These are
the group of R. Orbach in California University and the group of K.
Baberschke in Frei University in Berlin. The above groups have
published a whole series of papers on the EPR study of Gd$^{3+}$ ion
in both the normal and superconducting state of different
intermetallic compounds
\cite{Orbach1,Orbach2,Baberschke1,Baberschke2}. The detailed analysis of
the temperature dependencies of the linewidth and g-value performed
in these works allowed to get information concerning the behavior of
the electron magnetic susceptibility of a superconductor and the
peculiarities of its electronic structure. These studies have been
performed using the samples containing a small amount of magnetic
impurity in order to exclude the broadening of the resonance line
due to the spin-spin interaction. In principle, this information
could be obtained using the NMR experiments.

As to the conduction electron spin resonance (CESR) even in the
Abrikosov state due to the fast decrease of the elementary
excitations above the superconducting gap upon lowering the
temperature and large value of the spin-orbit interaction the
possibility for observation of CESR remained open for many years.

CESR was observed at the later stage by Yafet {\it et al.}
\cite{Schultz}. They reported the first observation of CESR in both
the normal and superconducting state of pure niobium. The resonance
displayed a g-value of 1.84$\pm$0.01 in both states. The linewidth
narrows considerably in the superconducting state. The authors
showed that this narrowing is a consequence of the coherence
effects.

The influence of paramagnetic impurities on the properties of
superconductors was one of the intensively investigated problems of
superconductivity physics in 1970ies. The observation of EPR of a
localized moment in type II superconductor added the EPR method to
the number of physical methods used to study this problem. Being a
direct detector of the formation of an electronic localized magnetic
state, this method yields information on a number of important
properties of a superconductor doped with a magnetic impurity. All
the possible applications of EPR to superconductors have not yet
been clarified completely, but the following can already be noted at
present. The EPR method makes it possible to measure directly the
exchange interaction of conduction electrons and the localized
states, to obtain detailed information on the spin scattering of the
conduction electrons individually for a given kind of impurity, to
investigate the collective spin-density oscillations produced at
sufficiently high concentrations of the paramagnetic impurity (the
"electron bottleneck" effect \cite{Hasegawa}) (see also APPENDIX
II). Finally, and apparently most significantly, EPR makes it
possible to investigate directly  the character and the strength of
the interactions between the electronic localized states in the
superconducting phase and consequently to investigate the problem of
the coexistence of magnetic order and superconductivity.

This paper presents the results obtained by the group of E. G.
Kharakhsh'yan concerning the first observation of EPR in type II
superconductor, observation of new type of an exchange interaction
between localized moments, magnetic ordering of impurities in the
superconducting state and, finally, the phase separation in
high-$T_c$ superconductors.

All EPR measurements were performed using an X-band spectrometer
equipped by the home-made helium cryostat (T=1.7$\div$4.2 K) and by
the flowing helium gas system (T=5$\div$300 K). First observation of
EPR was done using  spectrometer RE-1306 (Russia) and all other
measurements using EPR spectrometer B-ER 418$^s$ (Bruker).

\section{First observation of EPR in type II superconductor}

In this section we show that the observation of EPR of an electron
localized moment in a type II superconductor in its vortex state is
possible in principal. In contrast to NMR, for a long time the EPR
in superconductors was failed to be observed. Besides the
general difficulties in the observation of magnetic resonance
because of the Meissner effect and expected strong inhomogeneous
broadening of the line a short spin relaxation time appears to be an
additional barrier for the observation of the EPR. Even under
favorable conditions they do not exceed $10^{-8}$ s ($\sim$30 Oe).
These short relaxation times consequently entirely exclude the
possibility for using the pulse measurements and the method of
adiabatic cycling of magnetic field \cite{Slichter,Redfield} which
was successfully used in NMR. Another method which allows to
overcome the consequences of the Meissner effect and spacial
inhomogeneity of magnetic field  is the milling of the samples down
to the sizes comparable with the superconducting penetration depth
$\lambda$. However, in the case of EPR this method is noneffective
since the preparation of a homogeneous ensemble of particles with
the 100 {\AA} sizes and defined magnetic impurity concentration
represents an exceptionally complex technological task even for
single-component superconductors. (Here we discuss the preparation
of samples for the observation of EPR of the localized magnetic
states.)  For more complex multicomponent systems these difficulties
increase. Finally, the existence of magnetic impurities suppresses
superconductivity and decreases the critical parameters of
superconductor. The most heavily this effect concerns the transition
elements and weaker rare-earth elements. For the first observation
of EPR in a superconductor it was natural to choose as an
object of research a "bulk" (with dimensions $d>\lambda$) type II
superconductor doped by rare-earth impurity. At the beginning we
supposed that the search for the resonance signal should be
performed in the vortex state, when the magnetic field partially
penetrates into the sample and at the magnetic field close to the
upper critical field $H_{c2}$ in order to decrease the effect of
inhomogeneous broadening of the resonance line. Later on we obtained
that in order to observe the resonance line on the background of
baseline drift in a superconductor we should not be too close to the
superconducting critical field.

The first attempts to observe EPR of a localized moment have been
undertaken by us in the period from 1969 to 1971 for the metallic
lanthanum samples doped by the gadolinium impurity. From one side a
metallic lanthanum doped by rare earth magnetic impurity is a very
convenient matrix for the EPR study of rare-earth ions due to their
good solubility and possessing of tempered chemical activity.
Unfortunately, from another side it can be hardly obtained in a
single-phase. It is well known that the pure metallic
lanthanum crystallizes in two structural modifications: hcp with the
double axis ($\alpha$-La) and fcc ($\beta$-La). Both these
modifications are superconducting with the superconducting
transition temperatures $T_c$=4.9 and 6.0 K, correspondingly.
Lanthanum demonstrates all properties inherent in type II
superconductivity. Our measurements of the upper critical field for
$\beta$-La yield $dH_{c2}/dT\simeq$1.3 kOe/K at $T_c\simeq$6 K. In
principle, at $T\simeq$2 K we should have $H_{c2}\simeq$5 kOe. This
means that at $T$=2 K the resonance field $H_0\simeq$3350 Oe will be
smaller than $H_{c2}$ and we should be able to observe EPR of our
samples in the superconducting state. It is necessary to note that
such favorable situation should occur for pure La samples. Doping
the samples by Gd will decrease $T_c$. Our studies show that the
$T_c$-suppression by 1 at.\% of Gd $dT_c/dc\simeq$4 K/at.\%.

Nevertheless, we decided to perform the first measurements using the
dilute alloys of gadolinium in lanthanum. Altogether five samples
La$_{1-x}$Gd$_x$ with $x=5\cdot 10^{-4}\div 5\cdot 10^{-3}$ have
been studied. The samples were prepared by melting the constituents
under helium in a conventional induction furnace with subsequent
quenching in the cold helium steam after switching off the furnace.
As x-ray analysis has shown the quenching procedure allows to
prepare the samples with predominant content of the cubic
$\beta$-phase.

The EPR measurements were performed at the temperature range between
2 and 4.2 K. At all temperatures we observed a strong drift (almost
vertical) of the baseline of spectrometer which is caused by the
nonlinear field dependence of the surface impedance of a
superconductor. (EPR spectrometer records the first derivative of
the absorbed power on the dc magnetic field). In addition noises
arising due to the vortex motion when sweeping the dc magnetic field
further hamper the observation of the resonance. No resonance signal
was detected in such background. Thus, our first attempt failed.
Our analysis showed that the reason for this negative result was too
small value of $H_{c2}$ for the studied samples which only slightly
exceeds the resonance field. It was not enough to reach the
acceptable conditions for observation of EPR on the background of
the drift of the baseline. We also definitely observed that the
values of $H_{c2}$ determined from the drop in curve of the surface
impedance at $10^{10}$ Hz using the EPR spectrometer are noticeably
smaller than the values obtained using constant current. This drop
occurs at the dc magnetic field corresponding to equality
of the microwave field energy and the superconducting energy gap.
At small (compare to $H_{c2}$) magnetic fields the microwave energy
is much smaller than the superconducting energy gap and the upper
critical field determined from the drop of the surface impedance
coincides with the value determined from the changing
of the electrical resistivity using direct current.
At the same time with increasing the dc magnetic field
the superconducting gap decreases and the superconducting transition
detected via the microwave surface impedance occurs at smaller magnetic fields.

On the next step of our search we decided to repeat our attempt
using a superconductor with much higher critical parameters. That is
the La$_3$In intermetallic compound doped by gadolinium impurity.
Pure La$_3$In has $T_c$=9.1 K and $H_{c2}\simeq$70 kOe \cite{La3In}.

For our preliminary measurements \cite{Altshuler} we used the sample
La$_{2.99}$Gd$_{0.01}$In sample which was prepared by melting the
constituents in the tantalum crucibles under pure helium in a
conventional induction furnace. The sample for the measurements was cut
in a plate shape with dimensions 10x4x0.5 mm. The $T_c$-value for
our sample was of the order of 6 K. According to the data by Crow
{\it et al.} \cite{La3In} this allowed us to conclude that the
resonance field value $H_0$ lies well below the upper critical field
$H_{c2}$ and the observed EPR signal (Fig.~1)
\begin{figure}[t]
\centering{\includegraphics[width=0.75\columnwidth,angle=0,clip]{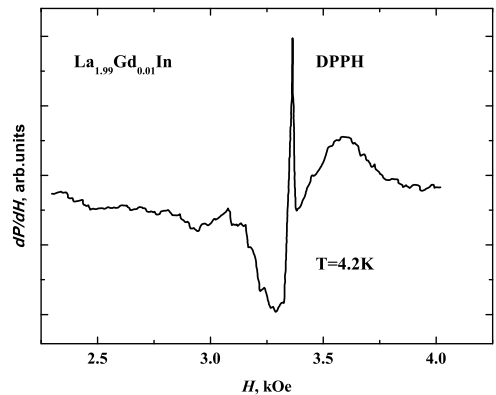}}
\caption{EPR spectrum for the La$_{2.99}$Gd$_{0.01}$In sample recorded
together with the signal of DPPH.}
\end{figure}
comes from the superconducting state of sample. In reality an EPR
signal was observed on the background of the field-dependent
baseline which was approximated by the straight line and subtracted
from the observed plot. EPR signal for our sample was recorded
together with the signal of the small amount of the free radical
DPPH (dyphenylpicrylhydrasyl) which was placed into the working
cavity at room temperature as a field label. As it is seen from
Fig.~1, the resonance signal has a typical asymmetric "metallic"
shape and consists of the mixture of the dispersion and absorption
curves. The pick-to-pick linewidth $\delta H=280\pm 20$ Oe and
$g=2.005\pm0.05$. It is interesting to note that the $\delta H$ and
$g$-values are almost equal to the same values for La$_{1-x}$Gd$_x$
at corresponding concentration of gadolinium at T=4.2 K in the
normal state. Thus, in this Section the possibility of observation
of EPR in type II superconductor was demonstrated. The continuation
of this study is presented in Section V.

\section{New type of exchange interaction}

Both in the first and in the subsequent papers on EPR of a localized
moment in superconductors, the investigations were carried out on
intermetallic compounds with small amount of the gadolinium
impurity. The choice of intermetallic compounds for the
investigations in
\cite{Altshuler,Orbach1,Orbach2,Baberschke1,Baberschke2} was
determined by the need to have relatively high critical parameters
for the samples doped with paramagnetic impurities. In the
investigations of such samples by the EPR method, however, there is
always the danger of the appearance of parasitic signals, which mask
the true effect. This may be caused either by the tend to oxidize
of these compounds, or by the difficulty of attaining the
necessary homogeneity and stoichiometry. It is therefore of interest
to carry out the EPR investigations on paramagnetic impurities in a
single-component type II superconductor. In this respect, metallic
lanthanum is very attractive, in view of the good solubilities of
rare-earth metals in it. The critical parameters of pure
single-component superconductors are usually lower than the
corresponding values for compounds. We met this situation when trying
to observe EPR of Gd$^{3+}$ ion in lanthanum and failed (see Section
II). Fortunately, this difficulty can be overcome choosing a
paramagnetic impurity with a large g-value, for example, Erbium with
$g\simeq 6.8$ corresponding to the resonance line field of 1000 Oe.
In addition Erbium  suppresses the superconductivity of the lanthanum
much less than gadolinium. Our measurements show that
$T_c$-suppression by erbium $dT_c/dc\simeq $0.4 K/at.\% Thus, the
measurements can be performed up to high concentrations of the
paramagnetic impurities, when the interaction between them becomes
substantial. As it was mentioned in Section II metallic lanthanum
crystallizes as a mixture of two modifications with hexagonal
($\alpha$-La) and face-centered cubic ($\beta$-La) lattices.
However, the choice of erbium as the paramagnetic impurity makes it
possible to disregard the presence of the hexagonal phase of the
lanthanum, since the strong anisotropy of the g-value of the
Er$^{3+}$ ion in a hexagonal crystal makes the EPR signal
unobservable in the $\alpha$-phase in a polycrystalline sample.

We present here the results of investigations of EPR of the erbium
localized moments in metallic lanthanum with relatively high
concentrations of the magnetic impurity. These results have been
published in \cite{NE1,Nottingham,NE2}.

\subsection{EPR measurements. Experimental results}

Besides the usual limit imposed on the real sensitivity of the EPR
spectrometer for a metal as a result of the skin
effect, in the superconducting state it is necessary to take into
account also the increased noise due to the motion of the vortices
and a strong dependence of the surface impedance of the
superconductor when sweeping the dc magnetic field. The latter
circumstance is particularly significant near the transition
temperature, where the EPR measurements are very difficult. The
minimum erbium concentration that could be observed in the
superconducting state was 0.5 at.\%. The maximum concentration was 6
at.\%.

{\it 1) Line shape.} The EPR line shape in the normal state had the
asymmetric shape usually observed for the bulk metallic samples and
was well described by a superposition of Lorentzian dispersion and
absorption curves. The linewidth (halfwidth of the absorption line
at the 1/2 amplitude) $\Delta H$ and the g-value were determined
using the standard procedure. Figures 2 and 3 show a plot of the
derivative of the resonant-absorption power of a sample in the
superconducting state.
\begin{figure}[t]
\centering{\includegraphics[width=0.7\columnwidth,angle=0,clip]{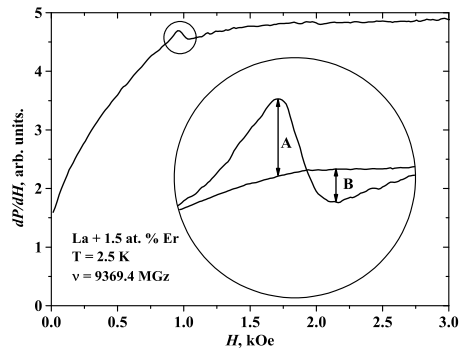}}
\caption{Plot of the EPR signal in the superconducting state.}
\end{figure}
\begin{figure}[t]
\centering{\includegraphics[width=0.7\columnwidth,angle=0,clip]{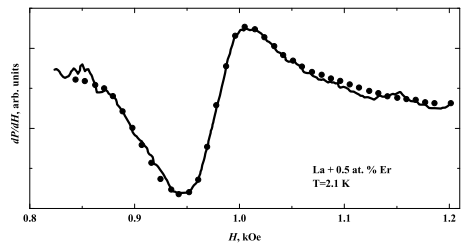}}
\caption{EPR signal in the
superconducting state. The points were calculated taking
into account the vortex lattice with $\Delta H$=35 Oe and
$H_v-H_c$=84 Oe.}
\end{figure}
Fig.~2 shows the background of the EPR signal at the temperature
well below $T_c$ at the resonance field. The line shape, especially
for small erbium concentrations (Fig.~3), differs substantially from
a pure Lorentzian, principally as a result of the inhomogeneous
distribution of the magnetic field inside the sample, caused by the
appearance of the vortices. The shape of the resonance signal in the
superconducting phase with allowance for the vortex lattice is given
in APPENDIX I. As seen from Fig.~3, the calculated curve describes
well the experimental spectrum. By reconciling the theoretical
curves with the experimental data it is possible to obtain the true
linewidth $\Delta H$ and the difference between the maximum $H_v$,
and the minimum $H_c$, fields in a sample that is in the mixed state
at $H_{c1}\leq H \leq H_{c2}$.

{\it 2) The g-value.} In the normal state, the g-value was equal to
6.80$\pm$0.05. The proximity of the observed value of the g-value to
the theoretical one for the doublet $\Gamma_7$, (for which g = 6.77
in a cubic field when account is taken of the intermediate coupling)
is evidence that the ground state in our case is the doublet
$\Gamma_7$, as follows also from our susceptibility measurements.
This is an unequivocal confirmation of the fact that the resonance
signal is due to the cubic $\beta$-phase of the lanthanum.

For the superconducting phase, when calculating the line shape by
the scheme described in APPENDIX I, it was assumed that the magnetic
field at the center of the vortex is equal to the external magnetic
field. Actually, as shown, for example, by NMR measurements of
single-crystal niobium samples of high quality, this field can
noticeably exceed the external field \cite{Rossier1}. Therefore the
values of the g-value, calculated from the spectrum distorted by the
vortex lattice, may differ from the true values. For this reason, we
do not analyze here the g-value in the superconducting region.

{\it 3) Temperature and concentration dependence of the linewidth.}
Figure 4 shows the dependence  $\Delta H(T)$ for several samples in
the studied temperature range, while Fig. 5 shows in greater detail
the low temperature part of this dependence for one of the samples.
\begin{figure}[t]
\centering{\includegraphics[width=0.75\columnwidth,angle=0,clip]{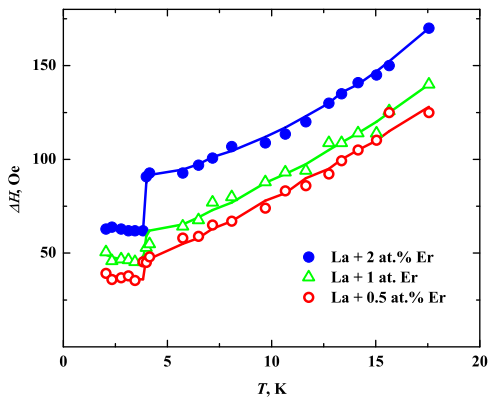}}
\caption{Temperature dependence of the linewidth for three samples. }
\end{figure}
\begin{figure}[t]
\centering{\includegraphics[width=0.9\columnwidth,angle=0,clip]{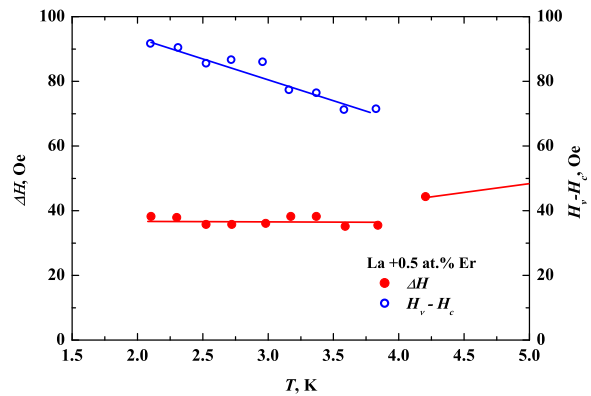}}
\caption{ Temperature dependence of the linewidth and of the difference between the
limiting values of the magnetic field in the vortex lattice for the
sample La+0.5 at.\% Er.}
\end{figure}
Fig. 6 shows the concentration dependence of $\Delta H$ in the
normal and superconducting states for all studied samples. These
figures allows us to conclude the following: a) at temperatures
below 14 K in the normal state, the temperature dependence of the
linewidth can be roughly approximated by the linear relation $\Delta H=a
+bT$; b) above 14 K, the temperature dependence of $\Delta H$ is no
longer linear; c) the coefficients $a$ and $b$ depend on the
concentration; d) upon transition into the superconducting state,
the linewidth decreases sharply, and the jump is proportional to the
erbium concentration; e) $\Delta H$ is constant in the
superconducting state.
\begin{figure}[t]
\centering{\includegraphics[width=0.75\columnwidth,angle=0,clip]{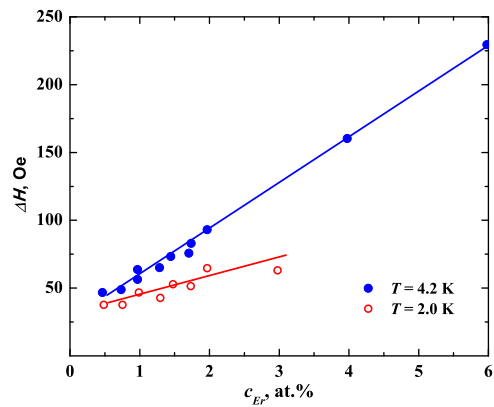}}
\caption{Concentration dependence of the EPR linewidth for the
La$_{1-x}$Er$_x$ samples.}
\end{figure}

\subsection{Discussion of results}

{\it 1) Normal state.} That part of the EPR linewidth which depends
linearly on the temperature is due to thermal fluctuations of the
exchange interaction of the localized $f$ electrons with the
conduction electrons (the Korringa mechanism \cite{Korringa}). If we
introduce for the doublet $\Gamma_7$, which is ground state of
Er$^{3+}$ in fcc La, an effective spin $S$ = 1/2, then the
Hamiltonian of the exchange interaction can be written in the form
$g(g_L - 1)JSs/g_L $ ($s$ is the spin of the conduction electrons),
and the exchange integral $J_{sf}$ is determined from the
temperature slope of the linewidth $b$ (see APPENDIX II). As seen
from Eq.(6), the measured quantity is the product of the exchange
integral by the density of states. If we use for the electron
density of state at the Fermi level the value $\rho (E_f)$ = 2
eV$^{-1}$atom$^{-1}$spin$^{-1}$ known from measurements of the heat
capacity then in the case of the sample with the smallest erbium
concentration, when the effect due to the collective oscillations of
the spin density of the localized moments and conduction electrons
are each significant, we obtain from Eq.(8) the value $J_{sf}$= 0.13
eV.

It is of interest to compare this quantity with the data obtained
from the dependence of $T_c$, on the concentration of the
paramagnetic impurity. The presented EPR and susceptibility results
show that at temperatures of the order of $T_c$, we can neglect the
scattering of the electrons by the excited levels of erbium (see
below). To obtain $J_{sf}$ in the presence of the crystal-field
effects we can therefore use the formula of Abrikosov and Gor'kov
\cite{AG} (see APPENDIX II),which we express in terms of the
effective spin. This yields a value $J_{sf}$= 0.04 eV, which differs
noticeably from the corresponding value obtained from EPR. The
observed discrepancy can be connected with the fact that in
lanthanum the conduction electrons belong mainly to the s and d
bands, and the contributions from these bands to the spin relaxation
rate (the Korringa and Overhauser relaxations) and to the heat
capacity may not be fully equivalent.

The proximity of the excited level to the ground level causes the
temperature dependence of the linewidth to deviate from linearity at
temperatures on the order of 14 K, as a result of the relaxation
process of the Orbach-Aminov type \cite{OA}. Allowance for this
mechanism leads to the temperature dependencies of the linewidth,
shown by the solid lines in Fig. 4. It is interesting to note that
all three set of crystal field parameters obtained in the
calculation of the magnetic susceptibility make the same
contribution to the relaxation.

The part $a$ of the EPR linewidth which does not depend on
temperature, can be ascribed to magnetic dipole-dipole interactions.
A contribution to the linewidth can also be caused by the distortion
of the spatial distribution of the charge density of the conduction
electrons, due to lattice defects. They lead to the appearance of a
low-symmetry contribution to the crystal field and cause the g-value
to be shifted as a result of mixing of excited states with the wave
functions of the doublet $\Gamma_7$. It is important to emphasize
here that whereas the scatter of the g-value is due to the mutual
influence of the paramagnetic impurities, the linewidth will
increase with increasing concentration, just as in the dipole-dipole
broadening mechanism. On the other hand, the mutual distortion of
the spin density of the conduction electrons by paramagnetic
impurities leads to the known indirect RKKY exchange \cite{RKKY},
which suppresses the indicated two-particle mechanisms of EPR line
broadening. The value of the RKKY exchange integral can be easily
estimated in the free-electron approximation (see, e.g.,
\cite{Taylor}):
\begin{equation}
J_{RKKY}={{9\pi Z^2 J_{sf}^2}\over{2E_f}}\phi (k_f R), \ \ \
\phi={{\sin x-x\cos x}\over{x^2}},
\end{equation}
where $Z$ is the number of conduction electrons per atom, $E_f$ is
the Fermi energy, and $k_f$ is the corresponding wave vector. To
this end it is necessary to calculate the value of the exchange
integral $J_{sf}$, using the density of states of the conduction
electrons $\rho (E_f)$= 0.33 ev$^{-1}$atom$^{-1}$spin$^{-1}$
calculated in the free-electron model for lanthanum. From the EPR
data we obtain $J_{sf}$= 0.08 eV, and from the dependence of $T_c$
on the erbium concentration we get $J_{sf}$=0.1 eV. The values of
$J_{sf}$ are in fair agreement. The use of $J_{sf}$=0.08 eV yields
for the nearest neighbors in the lattice the value $J_{RKKY}$ = 2 K,
which is much higher than the energy of the magnetic dipole-dipole
interaction (for which in our case the estimate yields - 0.05 K).
This means that an exchange narrowing of the EPR line takes place.

{\it 2) Superconducting state.} Upon transition to the
superconducting phase, an additional reason for the broadening of the
EPR line appears, because of the inhomogeneous distribution of the
magnetic field in the sample at $H_{cl}< H <H_{c2}$. The
experimental observed deviation of the shape of the EPR line from
Lorentzian in the phase transition agrees with the assumption that a
vortex structure is realized in our samples. The theoretical
EPR line shape can be calculated by introducing the distribution
function of the magnetic field in the superconductor (see APPENDIX
I). The value $H_v - H_c$= 70 Oe obtained by this method for a
sample containing 0.5 at. \% Er (see Fig.~3) can be compared with
the result obtained from measurements of the magnetic moment $M$ of
the superconductor. For a triangular vortex lattice we have $H_v -
H_c= - 1.46 \cdot  4\pi M$ \cite{Rossier1} which in our case yields
75 Oe, in good agreement with the value given above.

The most interesting feature of the results of the measurements in
the superconducting phase is the sharp narrowing of the resonance
line just below the superconducting transition temperature $T_c$.
This observation was in contradiction to the commonly accepted point
of view that the resonance line should broaden upon transition to
the superconducting state.   Let us consider the possible reasons
for this effect. As a result of the appearance of coherence effects
in the scattering of the conduction electrons the Korringa
relaxation rate \cite{Korringa} increases sharply near $T_c$ similar
to the case of the NMR relaxation \cite{Slichter}. At the same time,
according to Maki \cite{Maki} the rate of exchange scattering of the
conduction electrons $\delta_{ei}$ below $T_c$ (the Overhauser
relaxation) also increases, and the spin-orbit scattering (the
spin-lattice relaxation of the conduction electrons) decreases
sharply. This circumstance enhances the conditions of the "electron
bottleneck," and leads to a narrowing of the EPR line. This effect,
however, can explain only in part the observed narrowing of the
line, since the change of $\Delta H$ at high impurity concentrations
exceeds the contribution due to the Korringa mechanism which can be
obtained by the linear extrapolation of the $\Delta H(T)$ from $T_c$
to zero temperature.

Let us consider also another possible reason for increase of the
exchange narrowing of the EPR line in the superconductor.  The
change of the RKKY exchange interaction upon transition into the
superconducting state is determined by the correlations of the
conduction electrons with opposite spin orientations, as a result of
which the average spin susceptibility tends to zero. The local spin
polarization of the electrons near the paramagnetic impurity is then
canceled out as a result of the indicated correlations over much
larger distances, on the order of the coherence length
\cite{Anderson}. The corresponding change of the exchange integral
is
$$
\Delta J_{RKKY}={{9\pi^2Z^2J_{sf}^2}\over{E_f^2 (k_f R)^2}}k_b
T\sum_\omega{{\Delta^2\over{\Delta^2+\omega^2}}\sin^2(k_f R)\times}
$$
\begin{equation}
{\times\exp{ \left( -{{2\sqrt{\Delta^2+\omega^2}}\over{v_f}}\right)
R}},
\end{equation}
where $\hbar \omega= \pi k_B T(2n + 1)$, $\Delta$ is the order
parameter in the superconductor, and $v_f$ is the Fermi velocity on
the conduction electrons.

An estimate of this contribution by the method of moments shows that
the change of the exchange narrowing of the EPR line is small, since
the rate of reorientation of the localized spins depends on
$J_{RKKY}^2$. It must be noted, however, that the exchange field of
spins of one orientation on a paramagnetic ion in a superconductor,
at low impurity concentration, can greatly exceed the corresponding
value for the normal metal as a result of the long-range action.
We note in conclusion  that the taking into account properly both
mechanisms mentioned above, we can explain fully the observed
discontinuity of the linewidth.

Thus, to interpret the EPR line  narrowing effect one should suggest
the enhancement the conditions for the "electron bottleneck" and
approved theoretically the existence of the new type of the
long-ranged exchange interaction between localized paramagnetic ions
in the superconducting state. That is a kind of "super exchange"
interaction between localized ions, the Cooper pair assisted
(mediated) interaction  over the distance of hundreds, even
thousands angstroms.

\section{Magnetic ordering in the superconducting state}

In the previous Section it was shown that EPR provides a most
effective method for investigating the nature and strength of the
spin-spin interactions between localized electronic states in the
superconducting phase. It accordingly becomes possible to use the
EPR method to investigate one of the most controversial problems in
the physics of superconductors - that of the coexistence of magnetic
order and superconductivity. A large amount of papers on the experimental
study of this problem using various physical methods have been
published (see, e.g., the reviews by Maple \cite{Maple} and Roth
\cite{Roth}).

In the work reported in this Section we used the EPR method for the
first time to investigate the effect of magnetic ordering of
impurities in a superconductor. The superconducting compound
La$_3$In doped with Gd was chosen as the material for the
investigation. Just for this system we have observed EPR of a
localized moment in superconductor for the first time
\cite{Altshuler}.

For all the investigated samples in the normal state, the shape of
the EPR line of gadolinium  localized states had the asymmetric
shape usual for bulk metallic samples. The g-value and the linewidth
$\Delta H$ were determined using the standard technique. The width
of the EPR line in the superconducting state was determined taking
into account the vortex lattice (see APPENDIX I). The results of the
measurements of the resonance linewidth for the investigated
specimens are presented in Figs.~7, and 8.
\begin{figure}[t]
\centering{\includegraphics[width=0.75\columnwidth,angle=0,clip]{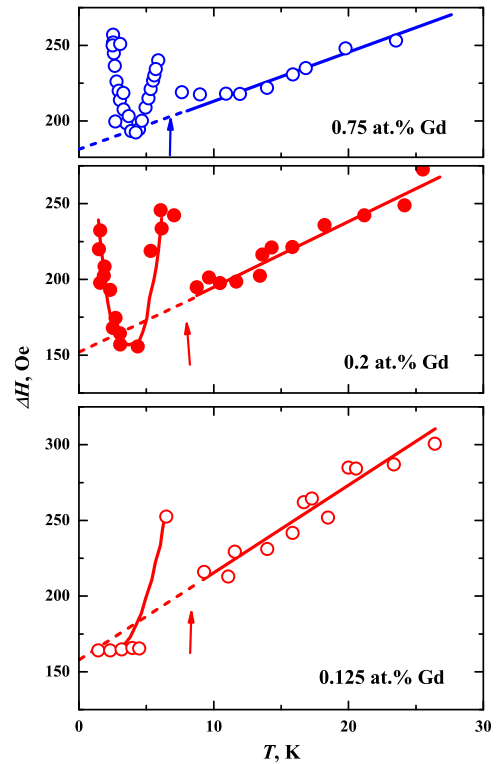}}
\caption{Temperature dependence of the linewidth for the
La$_{3-x}$Gd$_x$In samples containing different gadolinium impurity. The arrows show the transition temperature
to the superconducting state at the resonance field.}
\end{figure}
\begin{figure}[t]
\centering{\includegraphics[width=0.75\columnwidth,angle=0,clip]{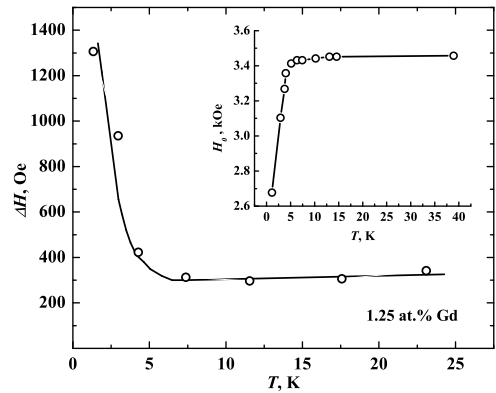}}
\caption{Temperature dependence of the linewidth $\Delta H$ and the
resonance field $H_0$ for the La$_{3-x}$Gd$_x$In sample. The solid curve in the main plot is the theory.}
\end{figure}

{\it The normal state.} The temperature dependence of the linewidth
in the normal state at high temperatures can be represented by a
linear function $\Delta H =a+bT$ in which the coefficients $a$ and
$b$ depend on the Gd concentration. That the temperature slope $b$
of the linewidth depends on the Gd concentration indicates that the
conditions for an electron bottleneck in the relaxation of localized
moments are satisfied (see APPENDIX II). In this case one can
determine important spin dynamic characteristics of the conduction
electrons from EPR data. From the temperature slope of the EPR
linewidth  using the value $\rho (E_f)$ = 1.7
eV$^{-1}$atom$^{-1}$spin$^{-1}$ which we determined from the
$dH_{c2}/dT$ data with the aid of the Gor'kov relation
\cite{Gorkov}, we obtain $\delta_{eL} =(8\cdot 10^{10})c$
sec$^{-1}$, $\delta_{ei}=(1 .7 \cdot 10^{13})c$ sec$^{-1}$, and
$J_{sf}$ =0.01 eV, where $c$ is the gadolinium concentration in
atomic percent.

One can also evaluate $\delta_{ei}$ from the dependence of the
critical temperature of the superconductor on the magnetic-impurity
concentration, using the well-known formula of Abrikosov and Gor'kov
\cite{AG} (APPENDIX II). The experimental value $dT_c /dc$ =3 K
yields $\delta_{ei}= (6.7 \cdot 10^{13})c$ sec$^{-1}$. The
difference between the values obtained for the exchange scattering
rate from the EPR data and from the $dT_c /dc$ data is evidently to
be attributed to the possibility that nonmagnetic scattering of
conduction electrons by gadolinium impurities may contribute
appreciably to the suppression of the superconductivity. In that
case the estimate of the exchange-scattering rate based on the
concentration dependence of the superconducting transition
temperature would be overestimated. Further, the possibility cannot
be entirely ruled out that the relaxation of localized gadolinium
moments in La$_3$In and the superconductivity of that compound may
be due to electrons of different bands. Moreover, it must be borne
in mind that the electron bottleneck effect arises only when the
relaxation of localized moments is due to $s$-band electrons, since
$d$ electrons have very short spin relaxation times.

The "residual" linewidth $a$ in the metal is determined by the
spin-spin interactions and the fine structure. Estimates show that
even in the absence of exchange narrowing, the contribution of
dipole-dipole interactions to the linewidth does not exceed 100 Oe
at the concentrations used in the present work. For the investigated
samples, therefore, the linewidth is evidently determined mainly by
the fine structure of the Gd$^{3+}$ ion and the inhomogeneities of
the crystal, whose contributions are partially averaged by spin
fluctuations induced by indirect exchange interactions.

As the temperature approaches the magnetic ordering point the
correlation range increases sharply under the action of the exchange
potential, and this leads to a corresponding increase in the
linewidth and to a shift of the resonance toward the weaker magnetic
fields. In the normal state, this effect is most clearly seen in the
sample containing the 1.25-at.\% Gd concentration (Fig.~8).

On the basis of all the experimental data (part of them are not
presented here), we can sketch the following qualitative picture of
the magnetic state of the impurities in the investigated compound.
At temperatures $T\leq 15$ K  strong antiferromagnetic correlations
between the neighboring spins were found from the magnetic
susceptibility data. As the temperature decreases further, the weak
exchange interactions, preferentially of ferromagnetic type at
distances $R\leq 3$ {\AA}, lead to "freezing" of the spin system
with the formation of a complex magnetic structure with a small
spontaneous moment.

It should be noted that the intensity of EPR signal deviating only
slightly from the Curie-Weiss law at low temperatures evidently give
us reason to assume that the impurity was uniformly distributed in
our specimens.

{\it The superconducting state.} The samples with gadolinium
concentrations of 0.125, 0.25, and 0.75 at.\% become superconducting
at temperatures in the range 6 $\div $8 K, the EPR linewidth first
increasing and then rapidly decreasing (Fig. 7). This behavior of
the linewidth is similar to the change in the nuclear relaxation
rate in a superconductor and is associated with the appearance of a
gap in the elementary excitation spectrum. As the temperature
decreases below 3 K the linewidth for the sample with the
0.125-at.\% Gd concentration does not change further, while those
for the samples with the 0.25- and 0.75-at.\% Gd concentrations
increase again.

As was noted above, for samples with low concentration of magnetic
impurity at low temperatures the contribution from antiferromagnetic
pairs to the spin-fluctuation frequency freezes out. It is therefore
natural to assume that the observed behavior of the linewidth for
the samples with Gd concentrations of 0.25 and 0.75 at.\% at
temperatures below 3 K is due to this effect.

In investigating the coexistence of magnetic order and
superconductivity it is important to answer the following question:
Do superconductivity and magnetic order coexist in the same volume,
or are the superconductive and magnetic phases spatially separated?
Since the basis for the conclusion that the gadolinium impurities
are magnetically ordered in the superconducting phase is the
characteristic behavior of the resonance line near the magnetic
ordering point, it is necessary to analyze the possibility that a
parasitic signal from extraneous inclusions in the normal state may
be superimposed on the EPR spectrum. The normal phase might be
produced either by a nonuniform distribution of the impurity
throughout the specimen or as a result of precipitation of a
nonsuperconducting modification of the investigated compound.

The degree of spatial uniformity of the impurity distribution can be
estimated from measurements of $T_c$ performed by the
ac-susceptibility method. This method provides information on the
variance of $T_c$ throughout the volume of the specimen, whereas
measurements of $T_c$ based on conductivity methods do not, since in
principle it is possible that normal-state inclusions may be shorted
out by superconducting regions having high critical temperatures. If
it is assumed that the fluctuation of the Gd concentration in the
sample is the main reason for the broadening of the superconducting
transition, the temperature dependencies obtained for the volume
that has passed into the superconducting state can be very well
described under the assumption that the impurity is distributed
normally in the specimen with a quite small standard deviation-for
example, with a standard deviation of $\sigma_c$ =0.08 \% for a
sample with $\sigma_c$ = 1.25 at.\% (i.e., 1.05 \%$\leq c\leq$ 1.45
\%, throughout 98 \% of the volume of the specimen). Moreover, the
EPR and magnetic susceptibility data on the sample containing 3
at.\% of Gd indicate that the distribution of the Gd$^{3+}$ ions is
uniform even on small scales of the order of the mean distance
between impurity atoms. As regards the presence in the sample of
extraneous modifications of the compound of lanthanum with indium,
it follows from the results of metallographic and x-ray studies that
such modifications amount to no more than 4 \%. The presence of 4 \%
or less of the normal phase in the investigated samples cannot
appreciably distort the observed temperature dependence of $\Delta
H$ in the superconductor since, as the results of special
measurements show, the parasitic signal would not be strong enough
to compete with the intense signal from the principal
superconducting phase. It should also be borne in mind that the
intensity of the EPR signal decreases very little (by 10-20 \%) at
the superconducting transition point.

Thus, our experimental data permit us to say with some confidence
that magnetic order and superconductivity coexist in a single phase
at temperatures below 3 K in the samples containing gadolinium
concentrations of 0.25 and 0.75 at.\%. Since it is reasonable to
assume that the temperature at which the line begins to broaden is
proportional to the true magnetic ordering point, we can assert that
the conditions for the appearance of magnetic order at least do not
become more stringent at the transition to the superconducting
phase; this becomes evident on comparing the data presented in Figs.
7 and 8. This is not even surprising, since theoretical calculations
\cite{NE2,Kochelaev} on transition to the superconducting state, the
indirect exchange interactions responsible for the magnetic order do
not undergo any substantial changes at distances shorter than the
coherence length, which, in the investigated samples, exceeds the
average distance between localized spins.

Thus, the use of the EPR method in the present work has made it
possible to detect the emergence of magnetic correlations between
spins in the superconducting phase itself; the results provide
grounds for concluding that in La$_3$In containing gadolinium
impurities, a magnetically ordered phase develops in the
superconducting state in practically the same way as it does in the
normal state.

This result was in a contradiction with Ginzburg's statement that
superconductivity and ferromagnetism are inconsistent. Indeed
superconductivity is the state in which pair of electrons with
opposite oriented spins form the Cooper pairs while ferromagnetism
is forming-up by the localized moments and polarized by them spins
of conduction electrons in one direction: these two phenomena are
antagonistic at any temperature. Observation of magnetic ordering in
the superconducting state by group of Kharakhash'yan was evidence
for the noncolinear magnetic ordering. It appeared that the magnetic
and the superconducting orderings can coexist if the mutual tuning
takes place.

\section{Phase separation in 1:2:3 high-$T_c$ system}

After the discovery of superconductivity above 30 K by Bednorz and
M\"{u}ller \cite{BedMul} in the La-Ba-Cu-O system, investigations on
the synthesis of new superconducting metal oxides were started. Soon
after, superconductivity at temperatures above the nitrogen boiling
point was discovered by Chu and collaborators \cite{Chu}  in
multiphase samples of Y-Ba-Cu-O. Subsequently it has been
established \cite{Cava} that the superconducting phase with $T_c$=90
K has the formula YBa$_2$Cu$_3$0$_{7-\delta}$. Among the several
striking features of these new superconducting systems, now commonly
named as 1:2:3-systems, their relation to magnetism is the most
surprising. Indeed, in x-ray homogeneous single phase
YBa$_2$Cu$_3$0$_{7-\delta}$, Curie-like temperature dependence of
magnetic susceptibility has been observed above $T_c$. Moreover, the
value of susceptibility varied over a wide range \cite{Kaji}.
Bearing in mind that superconductivity and magnetism are
antagonistic phenomena, it is surprising that $T_c$ is unchanged
when the Curie-like part of susceptibility is varied. The next
interesting feature of these systems is their indifference to the
magnetic state of rare-earth ion substituting for yttrium: $T_c$
over the rare-earth series is practically the same.

In this Section we report the results of our EPR studies of
1:2:3-system samples with the different oxygen content. Our results
have been published in Refs. \cite{NE5,NE6}. As a spin probe we used
the Gd$^{3+}$ ion in Y$_{1-x}$Gd$_x$Ba$_2$Cu$_3$0$_{7-\delta}$.

\subsection{Samples}

To obtain the Y$_{1-x}$Gd$_x$Ba$_2$Cu$_3$0$_{7-\delta}$ samples with
small gadolinium content, the Y$_{1-x}$Gd$_x$ alloy was first
prepared. Then it was dissolved in nitric acid, and the salt
solution that was obtained was evaporated and decomposed at the
temperatures of the order of 600$^o$ C. We believe that the
Y$_{1-x}$Gd$_x$O$_3$ oxide obtained was sufficiently homogeneous and
was used when synthesizing the samples. Changing the temperature of
synthesis, the annealing time, the rate of cooling, and the partial
oxygen pressure, we prepared the samples with a different degree of
the orthorhombic distortion.

The oxygen deficiency in YBa$_2$Cu$_3$0$_{7-\delta}$ samples was
determined using as a reference the correlation of the lattice
parameters with the oxygen content. It was found that the minimum
value of the oxygen deficiency for our samples corresponded to 0.15
and the maximum one to $\delta$ = 0.5. The oxygen concentration for
the samples containing gadolinium was evaluated by the same way.

To increase the signal-to-noise ratio, EPR measurements were
performed on powder samples prepared by grinding of pellets. The
powder with particle dimensions of the order of the single
crystallite sizes ($\sim 10\ \mu$m from optical microscopy) was
mixed with paraffin and placed into a quartz ampule.

\subsection{Resistivity}

For the samples with small oxygen deficiency, the linear temperature
dependence of the resistivity was observed above $T_c$. All the
samples with metallic behavior showed a sharp ($\Delta T_c\leq 2$ K)
transition from the normal to the superconducting state. When
lowering the oxygen content the temperature slope of the resistivity
became smaller. These samples had a broad transition to the
superconducting state. For the samples with a large oxygen
deficiency, semiconducting behavior of resistivity of the form $\rho
(T)\sim exp(A/T^\alpha)$ was observed. In particular for the samples
with $\delta$ = 0.5 the {\it A-} and $\alpha$-values were of the
order of $10^2$ K$^{1/2}$ and 1/2, respectively.

\subsection{The EPR of Gd$^{3+}$ ions in
Y$_{l-x}$Gd$_x$Ba$_2$Cu$_3$0$_{7-\delta}$}

The EPR measurements were performed at different temperatures in the
range between 1.5 and 300 K. Fig. 9 shows the EPR signals for
Y$_{0.99}$Gd$_{0.01}$Ba$_2$Cu$_3$0$_{7-\delta}$ samples with
$\delta$=0.15 and 0.5.
\begin{figure}[t]
\centering{\includegraphics[width=0.75\columnwidth,angle=0,clip]{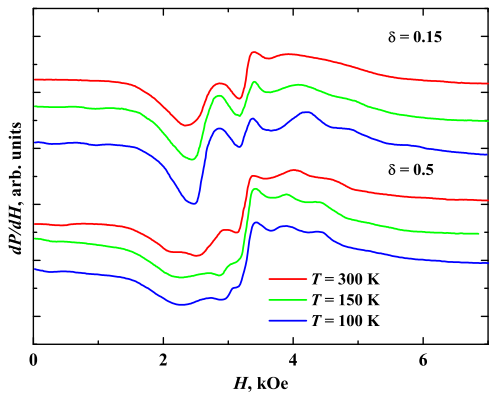}}
\caption{EPR signals for the Y$_{1-x}$Gd$_x$Ba$_2$Cu$_3$0$_{7-\delta}$ samples with different
oxygen deficiency at three different temperatures.}
\end{figure}
The observed fine structure of the EPR spectra is typical for powder
samples with a small gadolinium content. The single-line spectrum of
Gd$^{3+}$ ions was observed for the GdBa$_2$Cu$_3$0$_{7-\delta}$
samples with $\delta$= 0.15 and 0.5 (see Fig. 10).
\begin{figure}[t]
\centering{\includegraphics[width=0.75\columnwidth,angle=0,clip]{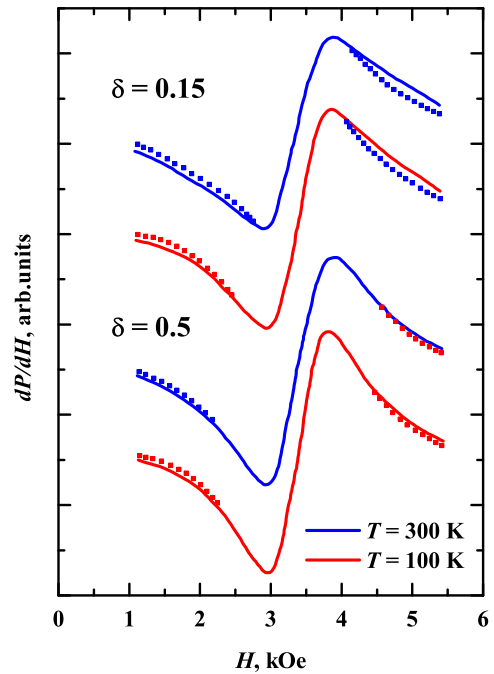}}
\caption{EPR signal for the GdBa$_2$Cu$_3$O$_{7-\delta}$ samples  with different
oxygen deficiency at two different temperatures. The solid line
is the experimental signal. Dotted line is the calculated one.}
\end{figure}
For these samples the fine structure is narrowed by the exchange
interaction between the gadolinium localized moments. Figure 11
shows the temperature dependence of the linewidth. In both samples,
at temperatures above 90 K the linewidth increased with the
temperature. For GdBa$_2$Cu$_3$0$_{7-\delta}$ sample near $T_c$= 83
K one can see a slight anomaly: broadening of the line immediately
below $T_c$ and then rapid narrowing. This behavior of the linewidth
is caused again by the appearance of the superconducting gap and
of the superconducting coherence effects. When lowering the
temperature below 50 K, the EPR signal begins to broaden and shift
to the low-field region. This broadening of the EPR line is caused
by magnetic ordering of Gd$^{3+}$ ions similar to the case discussed
in Section IV.

\subsection{Heterogeneity of System}

The temperature behavior of the Gd$^{3+}$ ion's EPR spectra for
Y$_{1-x}$Gd$_x$Ba$_2$Cu$_3$0$_{7-\delta}$ and
GdBa$_2$Cu$_3$0$_{7-\delta}$ samples with different oxygen
deficiency shows the Korringa-like behavior in the relaxation of
gadolinium localized moments. For the samples with $\delta$ = 0.15,
it is not surprising because the temperature dependence of
resistivity above $T_c$ has metallic character, but for the samples
with $\delta$ = 0.5, this result is unexpected. When we studied the
temperature dependence of resistivity, we obtained the
"semiconducting" behavior in the form $\rho (T)\sim
exp(A/T^\alpha)$.  Observation of such law with $\alpha$ = 1/4 for
the 1:2:3-system with large oxygen deficiency led Mei {\it at al.}
\cite{Mei} to conclusion that due to Anderson localization of
current carriers in such samples, hopping conductivity with varying
lengths takes place. Our data on EPR, however, rule out this
possibility. The point is that in the vicinity of Anderson
transition, where resistivity increases with the lowering the
temperature, localized moment's relaxation rate has also to increase
due to increasing of the correlation time of conduction electron's
spin fluctuations. We observed just the opposite effect: EPR
linewidth (the localized moment's relaxation) decreases with the
lowering the temperature.

Another possibility to understand the metallic character of EPR and
"semiconducting" behavior of resistivity for the samples with
$\delta$=0.5 is the supposition about the granularity of our
samples. In granular system the tunneling between metallic particles
occurs by the path where the combination of two factors -- transfer
integral $\exp{(-d/r_0)}$ and activation probability
$\exp{(-e^2/\varepsilon dk_B T)}$ -- is most favorable when varying
the distance $d$ between granules. Here $r_0$ is the radius of
localization and $\varepsilon$ is the dielectric permeability of
nonmetallic regions. As a result, for resistivity, one can obtain
the law $\rho (T)\sim exp(A/T^{1/2})$), with $A=(4e^2 r_0
\varepsilon k_B)^{1/2}$, which is close to the law observed in our
experiments.

So the different behaviors of the resistivity and metallic character
of EPR may be consistent with each other if one assumes that samples
under study are heterogeneous and consist of metallic and dielectric
phases \cite{NE5,NE6} One may think that for the sample with
$\delta$ = 0.15, there is percolation, while for $\delta$ = 0.5
there is not. More unequivocal conclusions may be derived from the
analysis of the EPR lineshape for the samples of
GdBa$_2$Cu$_3$0$_{7-\delta}$. The lineshape of the Gd$^{3+}$ EPR
signal (Fig. 10) shows a noticeable deviation from a Lorentzian,
especially for the GdBa$_2$Cu$_3$0$_{6.85}$ sample. We suppose that
the observed EPR signal consists of two signals with nearly equal g
values ($g\simeq 1.99)$ but with different linewidths. The best
simulation of the observed EPR spectra for any temperatures see
(Fig.~10) was obtained by summing up two Lorentz lines with the
following linewidth values. For the first line, the linewidth
linearly depends on the temperature in accordance with the $\Delta H
= a + bT$ law, where $a$ = 580 Oe and $b$ = 0.9 0e/K. For the second
one, the linewidth does not depend on the temperature and is equal
to 1000 Oe. We suppose that these signals are determined by metallic
and dielectric components, respectively. The pronounced nonlinearity
of the observed temperature dependence of $\Delta H$ in the normal
state (Fig. 11) proves to be the result of interference of these two
\begin{figure}[t]
\centering{\includegraphics[width=0.75\columnwidth,angle=0,clip]{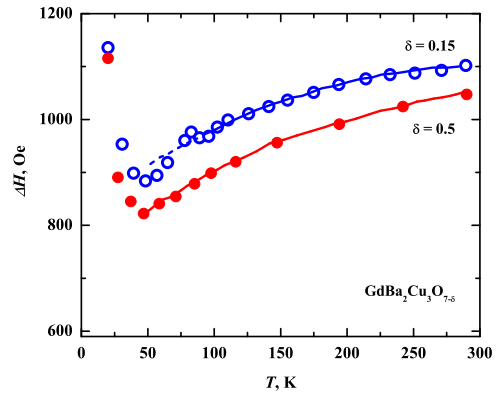}}
\caption{Linewidth as a function of temperature for two
GdBa$_2$Cu$_3$O$_{7-\delta}$ samples. Solid lines
are the resulting linewidths of simulating spectra obtained as a sum
of two Lorentz lines with $\Delta H=(580+0.9 T)$ Oe and $\Delta
H$=1000 Oe.}
\end{figure}
signals. Analysis shows that for the GdBa$_2$Cu$_3$0$_{6.5}$ sample,
the contributions from the dielectric and metallic regions to the
EPR signal are equal. The lack of dispersion contribution to the EPR
signal indicates that the thickness of the metallic regions in this
sample is much smaller than the skin depth. That is why the fraction
of metal may be estimated as 50\%. For the GdBa$_2$Cu$_3$0$_{6.85}$
sample, the integral intensity of the EPR signal was three times
smaller than for the GdBa$_2$Cu$_3$0$_{6.5}$ sample. Since the
number of Gd$^{3+}$ spins is equal for both samples, the loss of the
signal intensity may be caused only by the skin effect. If the metal
portion is designated by $y$ and the EPR signal intensity in the
absence of the skin effect is assumed to be equal to one, then, for
GdBa$_2$Cu$_3$0$_{6.85}$, it may be written that $\beta y+(1-y)$ =
1/3 (here $\beta$ is the relative portion of metal which forms EPR
signal). Taking into account that contribution of dielectric and
metallic regions to the simulated spectrum relate to each other as
7:3, it may be written that $(1-y)/\beta y$ = 7/3. From these two
relations, $y$ =0.77 and $\beta$= 0.13 is found. Thus the fraction
of metal in this sample may be estimated as 77\%. At the same time,
for this sample there is a loss of about 87\% of the metallic EPR
signal intensity due to the skin effect. However, when simulating
the EPR spectrum, it was necessary to admix -20\% dispersion signal
to absorption signal in contrast to the situation of bulk metallic
samples, where the equal mixture of dispersion and absorption
signals is as expected \cite{Blombergen}. This probably suggests
that in the studied samples, metallic regions form a net in which no
less than 10\% of the metal volume has characteristic dimensions
considerably smaller than the skin depth.

\subsection{Discussion}

Our data on EPR of Gd$^{3+}$ions give evidence that the samples
studied with 1:2:3-structure are heterogeneous and consist of
metallic and dielectric regions. The variation of the volume ratio
of metallic and dielectric phases established in this study allows
us to suppose that in metallic regions the oxygen deficiency
$\delta$ is close to zero and in dielectric ones is close to 1.
Proceeding from the above, many peculiarities of the 1:2:3-system
may be understood. In particular, small critical current values,
which are probably limited by availability of weak links between
metallic regions, become clear. Different behavior of the
resistivity with temperature which has been obtained for samples
with different oxygen content may be also comprehended in the
framework of this paper.

Our data seem to be consistent with results obtained by other
methods. Thus electron microscopy investigations \cite{Tranquada}
utilizing microdiffraction and selected area diffraction on isolated
single-crystal grains of the orthorhombic phase of
YBa$_2$Cu$_3$0$_{7-\delta}$ revealed local variations in
orthorhombic parameter $\Delta a/a = 2(b-a)/(b + a)$. It was found
that the microscopic values of $\Delta a/a$ varied from zero to more
than twice the macroscopic value when scanning the surface of the
sample by an electron beam. A pseudobinary phase diagram was
proposed by Caponi {\it et al.} \cite{Caponi} based on the $\log
{P_{O_2}}-1/T$ plots, quenching investigation, and high-temperature
x-ray diffraction analysis. Different regions of the phase diagram
were proposed to correspond to the first orthorhombic phase ($O_1$),
the second orthorhombic phase ($O_{II}$), the tetragonal phase $T$
and coexistence of two phases. In Ref. \cite{Kuiper} the electron
micrograph study gave the possibility of discussing the mechanism
for the $O_I\rightarrow O_{II}\rightarrow T$ phase transformations
caused by the removal of oxygen atoms from the samples. If one
supposes that the tetragonal $T$ phase is a dielectric and the
orthorhombic $O_I$ phase is a metal, then the paramagnetic regions
forming the copper-EPR may be referred to the orthorhombic $O_{II}$
phase.

All of the results obtained in Refs. \cite{Tranquada,Caponi,Kuiper}
and in the present Section are consistent with the sample being in
spinodal decomposition region as predicted by Khachaturyan et al.
\cite{Khachaturyan1,Khachaturyan2}. They pointed out that the
nonstoichiometric compound YBa$_2$Cu$_3$0$_{7-\delta}$ is
thermodynamically unstable at low temperature. They postulated that
under ordinary cooling conditions, the phases have no time to reach
equilibrium and thus constitute a spinodally decomposed sample with
a microscopic distribution of oxygen vacancy ordering. Thus, the
slight variations in preparation conditions may be the reason for
different physical properties of the samples obtained.

In conclusion, the EPR method happened to be rather informative in
studying unusual properties of high-$T_c$ superconductivity. In
spite of high values of the superconducting transition temperature
$T_c$ the critical current values which destroy superconductivity
were not too much.  The temperature behavior of the Gd$^{3+}$ ions
spectra for the samples with $\delta$=0.15 and 0.5 demonstrates the
"metallic" character of the EPR while the electric resistivity for
the sample the with $\delta$=0.5 manifests the "semiconducting"
behavior. These observations lead us to the supposition about
granularity of samples. In granular system due to the tunneling of
electrons between metallic particles resistivity reveals the
"semiconductive" feature as it was observed in our experiments. So
the different behaviors of the resistivity and metallic character of
EPR for the samples with $\delta$=0.15 and 0.5 may be consistent
with each other if one assumes that samples under study are
heterogeneous and consist of metallic and dielectric components.

\section{Concluding remarks}

This historical review of the past EPR research on superconductors
in Kazan would not be complete without making a link to the present
time. In particular, early EPR experiments on the high temperature
superconductor YBCO is worth to assess in the current perspective.
These works have been performed soon after the advent of high-$T_c$
superconductivity (HTSC) and since then an enormous progress in the
understanding of the normal state and superconducting properties of
the cuprates has been achieved. With this accumulated knowledge that
early conclusion on the intrinsic heterogeneity of YBCO gained from
the EPR data would sound not really surprising. It is well
established now that the physics of HTSC is driven by the interplay
of spin, charge and lattice interactions which give rise to novel
electronic and magnetic phases and may even be a cause  of the HTSC
(see, e.g., popular reviews
\cite{Orenstein,Anderson1,Dagotto,Bonn}). The entanglement and
competition of different states manifest in such exotic phenomena as
the "fractionalization" of an electron in terms of the separation of
spin and charge and the emergence of stripe phases where magnetic
insulating phase is separated by the (super)conducting phase on a
nanoscale. But at the time of our EPR experiments these
"revolutionary" concepts that contradicted the wisdom of
conventional uncorrelated metals and superconductors  have not even
been proposed. In that sense we were very fortunate to obtain one of
the first experimental evidences of the intrinsic multiphase
character of YBCO where the simultaneous occurrence of magnetic and
superconducting regions indeed turns out to be an inherent property
of HTCS cuprates.

Experimental EPR works beginning from the mid of
the 90-ths of the past century dedicated to a further exploration
of an interplay of superconductivity and magnetism in the cuprates
are nicely reviewed in the paper by B. Elschner and A. Loidl \cite{Elschner}.
The readers interested in more details are refereed to this article.
Here, it is appropriate to mention a few of those works which have attracted
that time a lot of interest in the condensed matter community.
Introducing local EPR spin probes in superconducting cuprates turns out
to be a fruitful approach.  As an example, the group around A. Janossy
has employed high-field EPR on Gd doped YBCO to study the properties
of the evading pseudogap phase \cite{Janossy97,Janossy00}.
The group around V. Kataev has focused on Gd-EPR studies
of the antiferromagnetic dynamics in the stripe phase of
(La,Eu)$_2$SrCuO$_4$ \cite{Kataev97,Kataev98}.
The group around B. Elschner has used alternatively Mn$^{2+}$ ions as spin probes
imbedded directly in the CuO$_2$ planes to study the spin correlations
in La$_{2-x}$Sr$_x$CuO$_4$  \cite{Elschner94}. This group has even reported
an intrinsic EPR signal in La$_{2-x}$Sr$_x$CuO$_4$  which has been attributed
to the formation of a three-spin polaron, consisting of two Cu$^{2+}$ ions
and one p hole \cite{Elschner97}.  Besides cuprates, other "less correlated"
superconductors have been addressed with EPR in the near past. Particularly
remarkable was an observation of conduction electron spin resonance
in superconducting MgB$_2$ \cite{Janossy01}. Finally, soon after the discovery
of a new class of high temperature superconductors on the basis of iron pnictides
in 2008 \cite{Kamihara08}, the EPR method has begun to make valuable contributions
to the understanding of the physics of these novel materials.
One can mention a number of recent interesting works addressing
the normal and superconducting properties of iron arsenides by Eu-EPR
in the so-called 122-family \cite{Pascher10,Dengler12} and by Gd-EPR
in the 1111-family \cite{Alfonsov11,Alfonsov12}.

Other results of our study performed from 1972 to 1977 also should
be mentioned. They are the following.

1. We found an experimental evidence for existence of the new type
of the indirect exchange interaction between the localized moments
through the electrons constituting the Cooper pairs.

This superconducting indirect exchange interaction has been
predicted theoretically by Anderson and Suhl \cite{Anderson}. The
physical origin of this interaction is the following. In the normal
state the main exchange between the rare-earth localized moments in
metals is the Ruderman-Kittel-Kasuja-Yosida indirect exchange
interaction \cite{RKKY} via the conduction electrons. It means that
the localized moment due to the {\it sf}-interaction polarizes the
spin of conduction electron. This spin polarization oscillates in
space with changing its sign. Spin of another localized moment via
the same {\it sf}-interaction feels this polarization. The sum of
this indirect exchange over the lattice gives a possibility to
predict the type of magnetic ordering which would occur at low
temperatures. At high concentrations of magnetic impurities the
ferromagnetic or antiferromagnetic orders may occur. At small
impurity content very often a spin glass state is realized. In the
superconducting state the situation is rather different. As in the
case of the normal state the spin of the localized moment polarizes
the spin of electron. Another electron constituting the Cooper pair
at the distance of the order of the superconducting coherence length
$\xi$ (the Cooper pair size) due to the {\it sf}-exchange
interaction polarizes the spin of another localized moment in
opposite direction. Thus the range of the indirect exchange
interaction increases from the mean-free path $l$ of conduction
electrons ($\sim$ 20 {\AA}) in the normal state up to the
superconducting coherence length $\xi$ (for the LaEr dilute alloys
$l<<\xi$) in the superconducting state. This new superconducting
exchange due to its antiferromagnetic origin may lead to a new type
of the magnetic ordering called by Anderson and Suhl \cite{Anderson}
as a cryptoferromagnetic state. In the limiting case it is a
small-scale domain state with a period of the order of the coherence
length \cite{BB}.

2. We also proved that the superconductivity and magnetic ordering
of impurity may coexist.

Now it is well established that in order to coexist
superconductivity and ferromagnetism should mutually adjust to each
other. For example, the complex study of some ternary borides (see,
e.g., review by Buzdin and Bulaevskii \cite{BB} and references
therein) gave a possibility to conclude that at some temperature
range cryptoferromagnetism is realized in the superconducting state.
When lowering the temperature further, ferromagnetism arises and
superconductivity disappears.  To determine which state is
preferable: the superconducting/cryptoferromagnetic or the
normal/ferromagnetic, it is necessary to compare the free energies
of these two possibilities.

More exciting examples of mutual tuning of superconductivity and
ferromagnetism were found in the thin film heterostructures
consisting of superconducting and ferromagnetic layers. In these
systems the superconductivity and ferromagnetism are separated in
space. The properties of the superconducting and the ferromagnetic
layers and coupling between the layers can be controlled and changed
independently. This opens the possibility to observe much more
interesting phenomena than in alloys and intermetallic compounds. In
particular, one of them is the evidence for arising the
cryptoferromagnetism in the Pd$_{1-x}$Fe$_x$ layer at its small
thicknesses in the S/F bilayer. This result was obtained in
epitaxial V/Pd$_{1-x}$Fe$_x$ using the FMR technique
\cite{Garifullin}. Another one, the so-called spin screening effect
when the spin polarization of conduction electrons in the
ferromagnetic layer leads to the polarization of electron spins in
the superconducting layer with opposite sign at the distance of the
order of the superconducting coherence length from the F/S
interface, was found recently \cite{Salikhov1,Salikhov2}. The scale
of this effect is the coherence length because it is also caused by
the superconducting correlations.

The author is grateful to Dr. Vladislav Kataev for helpful
discussion. Further we thank Dr. Aidar Validov for technical support
in the spectra restoration.

\newpage

APPENDIX I\\
{\it EPR line shape in a superconductor}\\
\bigskip
In the previous Section we present the experimental evidence for the
possibility to observe EPR of a localized moment in the
superconducting state. Surprisingly the observed linewidth did not
exceed its value in the normal state. This means that in order to
understand the reason for unchanged linewidth upon transition from
the normal to the mixed superconducting state it is necessary to
calculate line shape in the Abrikosov \cite{Abrikosov} vortex state
taking into account the magnetic field distribution \cite{note}.
This mixed state is characterized by the presence of an ordered
lattice of tubes of magnetic flux, or vortices, surrounded and
maintained by superconducting currents. The magnetic field decreases
from a maximum value at the center of an isolated vortex over a
characteristic distance $\lambda $, the London penetration depth.
The superconducting order parameter grows from zero at the center of
the vortex over a distance of the order of $\xi (T)$, the coherence
length, with $\xi (T) < \lambda (T)$. For applied fields far from
both $H_{c1}$ and $H_{c2}$ (the lower and upper critical fields,
respectively) the vortex density is not too great. Experiments by
Redfield \cite{Redfield} in this region confirmed the ordered
arrangement of the vortices, and verified that the lattice symmetry
was triangular.

The distribution of the probabilities of encountering a given
magnetic field in a triangular vortex lattice can be approximated by
the analytic function \cite{Rossier1}
\begin{equation}
f(x)= \left\{ \begin{array}{lcl}
0.837-0.500\ln{(-x)},&\mbox{ $0.08<x<0$} \\
0.236-0.576\ln{x},   &\mbox{ $0<x<0.92$}
\end{array} \right.
\end{equation}
Here $x = (H - H_s)/(Hv - H_c)$, $H_v$ and $H_c$ are the maximum and
minimum fields in the lattice, $H_s$ is the field at the saddle
point of the unit cell of the vortex lattice. If the shape of the
homogeneously broadened EPR line in metal is determined by the
Lorentz line
\begin{equation}
I_{norm}(H)={{1}\over{\pi}}{{{\Delta H}+H}\over{(\Delta H)^2+H^2}}
\end{equation}
where $\Delta H$ is the half-width of the line at half-height and
$H$ is the magnetic field relative to the center, then the EPR line
shape with allowance for the broadening due to the vortex lattice is
determined by the convolution
\begin{equation}
I_{sup}(H)=\int{I_{norm}(H_1-H)f\left(
{{H_1-H_c}\over{H_v-H_c}}\right)dH_1}
\end{equation}

The observed behavior of the spectrum is qualitatively consistent
with the absence of pronounced broadening in the superconducting
state due to the vortex structure at certain values of parameters.
The average internal field is smaller than in the normal state; that
is, a part of the flux is excluded. Increasing the amplitude of the
high-field wing of the resonance line is due to the discontinuity at
the maximum field $H_v$.\\

\bigskip
APPENDIX II\\
{\it Electron bottleneck and $T_c$-suppression by magnetic impurity}\\

For nonmagnetic metals NMR relaxation rate always complies to the
Korringa law \cite{Korringa}. That is the linear dependence of the
spin lattice relaxation rate on the temperature. In contrast to NMR
in EPR in metals conduction electrons cannot generally be considered
to be in equilibrium and this leads to some complications in the
theory. The best known of these is the "electron bottleneck" effect.
If cross relaxation between the localized moments and conduction
electrons is rapid, spin angular momentum transferred to the
conduction electrons can be transferred back to the localized
moments before it has time to decay to the lattice. The apparent
rate of relaxation of the local moments is then determined by the
conduction electron's spin lattice relaxation (see Fig. 12).
\begin{figure}[t]
\centering{\includegraphics[width=0.75\columnwidth,angle=0,clip]{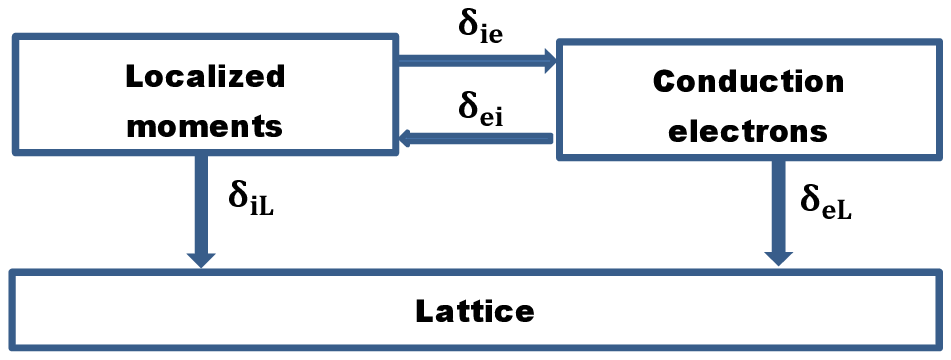}}
\caption{The relaxation paths involved in the EPR bottleneck. The
rates $\delta_{ie}$ and $\delta_{ei}$ result from the exchange
interaction; these rates denote transfer of magnetization from and to the
localized moments and conduction electron subsystems. }
\end{figure}
The bottleneck affects the position as well as the width of the
resonance line.

There are two regimes for EPR. First, isothermal regime
($\delta_{ei}<<\delta_{eL}$). The thermal broadening of the EPR
linewidth is determined by the Korringa law

\begin{equation}
\delta_{ie}^{Kor}={{\pi}\over{\hbar}}[J_{sf} \rho (E_f)]^2 k_B T,
\end{equation}
and the $g$-shift
\begin{equation}
\Delta g_{max}=J_{sf}\rho (E_f),
\end{equation}
where $\hbar$ is the Plank constant, and $\rho (E_f)$ is the density
of states of the conduction electrons with given spin orientation on
the Fermi surface, $J_{sf}$ is the $sf$-exchange integral between
localized moments and conduction electrons.

Experimentally one observes in addition a residual linewidth $a$
\begin{equation}
\Delta H=a+bT,\ \ \ b={{\hbar \delta_{ie}^{Kor}}\over{g\mu_B T}}.
\end{equation}
Second, bottleneck regime ($\delta_{ei}\geq\delta_{eL}$).

Very often for high concentration of impurity $\delta_{ei}$ may be
of the same order of magnitude as $\delta_{eL}$. Then the subsystem
of the conduction electrons is not longer in thermal equilibrium
with lattice. The effective relaxation rate $\delta_{ie}$ is
reduced. This was first calculated by Hasegawa \cite{Hasegawa}.

Assuming $g_i \simeq g_e$, the theory yields
\begin{equation}
\delta_{ie}={{\delta_{eL}}\over{\delta_{ei}+\delta_{eL}}}\delta_{ie}^{Kor},
\ \ b={{\hbar}\over{g
\mu_B}}\left({{\delta_{eL}\over{\delta_{ei}+\delta_el}}}\right)
{{\delta_{ie}^{Kor}\over{T}}}.
\end{equation}
\begin{equation}
\Delta g=\left({{\delta
eL}\over{\delta_{ei}+\delta_{eL}}}\right)\Delta g_{max}.
\end{equation}
Thus $\delta_{ie}^{Kor}$ corresponds to the "Korringa rate" and
$\delta_{ei}$ to the "Overhauser rate" which can be also obtained
from "detailed balance" to susceptibility of the two subsystems
$\chi_i/\chi_e=\delta_{ei}/\delta_{ie}$. Thus
\begin{equation}
\delta_{ei}={{2\pi}\over{\hbar}}\rho (E_f)J_{sf}^2 S(S+1)c.
\end{equation}

From another side, $\delta_{ei}$, is the controlling parameter for
pair-breaking in the superconducting state. The theory by
Abrikosov-Gor'kov \cite{AG} well describes the $T_c$-suppression by
magnetic impurity.
\begin{equation}
\ln (T_c/T_{c0}=\psi ({{1}\over {2}})-\psi \left( {{1}\over
{2}}+0.14{{\alpha}\over{\alpha_{cr}}} {{T_{c0}}\over{T_c}}\right).
\end{equation}
The pair breaking parameter $\alpha$ and its critical value
$\alpha_{cr}$, where superconductivity is completely destroyed, are
given by the theory in the first Born approximation
\begin{equation}
4\hbar \alpha_{cr}=k_BT_{c0}/\gamma;
\end{equation}
\begin{equation}
4\hbar \alpha_{cr}=c \rho(E_f)J_{sf}^2 S(S+1).
\end{equation}
This equation yields
\begin{equation}
\left|{{dT_c}\over{dc}}\right|={{3\hbar\pi}\over{16k_B}}{{d\delta_{ei}}\over{dc}}.
\end{equation}
This means that
\begin{equation}
\left|{{dT_c}\over{dc}}\right|=-{{\pi^2}\over{8k_B}} c\rho
(E_f)J_{sf}^2S(S+1).
\end{equation}


\newpage

\begin{thebibliography}{99}

\bibitem{BCS} J. Bardeen, L. N. Cooper, J. R. Schriffer, Phys.~Rev. {\bf 106},
162 (1957); {\bf 108}, 1175 (1957).

\bibitem{Slichter} L. C. Hebel and C. P. Slichter, Phys. Rev. {\bf
107}, 901 (1957).

\bibitem{Weger} M. Weger, {\it  Preprint} (1972).

\bibitem{Redfield} W. Fite, A. G. Redfield, Phys. Rev. {\bf
162}, 358 (1967).

\bibitem{Abrikosov} A. A. Abrikosov, Zh.~Eksper.~Teor.~Fiz.
{\bf 32}, 1442 (1957) [Soviet Physics JETP {\bf 5}, 1174
(1957)].

\bibitem{Altshuler} T.S.Al'tshuler, I.A.Garifullin, and E.G.Kharakhash'yan,
Fiz. Tverd. Tela {\bf 14}, 263 (1972)  [Sov.Phys. - Solid
State {\bf 14}, 213 (1972)].

\bibitem{Orbach1} C. Rettori, D. Davidov, P. Chaikin, R. Orbach,
Pys. Rev. Lett. {\bf 30}, 437 (1973).

\bibitem{Orbach2} C. Rettori, D. Davidov, R. Orbach, E. P. Chock, B.
Ricks, Phys. Rev. B {\bf 7}, 1 (1973).

\bibitem{Baberschke1} U. Engel, K. Baberschke, G. Koopman, S.
H\"{u}fner, Sol. State. Comm. {\bf 12}, 977 (1973).

\bibitem{Baberschke2}  K. Baberschke, U. Engel,S.
H\"{u}fner, Sol. State. Comm. {\bf 15}, 1101 (1974).

\bibitem{Schultz} Y. Yafet, D. C. Vier and S. Schultz, J. Appl.
Phys. {\bf 55}, 2022 (1984).

\bibitem{Hasegawa} H. Hasegawa, Progr. Theor. Phys. {\bf 27},
483 (1959).

\bibitem{La3In} J. E. Crow, R. P. Guertin, K. D. Parks,
Phys.~Rev.~Lett. {\bf 19}, 77 (1967).

\bibitem{NE1} N. E. Alekseevskii, I. A. Garifullin, B. I. Kochelaev, and
E. G. Kharakhash'yan, Pis'ma v Zh. Eksp. i Teor. Fiziki {\bf
18}, 323 (1973) [Sov.Phys. - JETP Lett. {\bf 18}, 189 (1973)].

\bibitem{Nottingham} B. I. Kochelaev, E. G. Kharakhash'yan, I. A. Garifullin, N.
E. Alekseevskii, Proceedings of 18th AMPERE Congress,
Nottingham, 1974, p. 23.

\bibitem{NE2} N. E. Alekseevskii, I. A. Garifullin, B. I. Kochelaev, and
E. G. Kharakhash'yan, Zh. Eksp. Teor. Fiz. {\bf 72}, 1523
(1977) [Sov. Phys. JETP {\bf 45}, 799 (1977)].

\bibitem{Rossier1} D. Rossier and D.~E.~MacLaughlin,
Phys.~kondens.~Materie {\bf 11}, 66 (1970).

\bibitem{Korringa} J. Korringa, Physica (Utrecht) {\bf 16}, 601
(1950).

\bibitem{AG} A. A. Abrikosov and L. P. Gor'kov, Zh. Eksp.
Teor. Fiz. {\bf 42}, 1242 (1961) [Sov. Phys. JETP {\bf 16},
879 (1963)].

\bibitem{OA} R. Orbach, Proc. R. Soc., London, Ser. A {\bf
264}, 458, 485 (1961); L. K. Aminov, Zh. Eksp. Teor. Fiz. {\bf
42}, 783 (1962) [Sov. Phys. JETP {\bf 15}, 547 (1962)].

\bibitem{RKKY} M.A. Ruderman and C. Kittel, Phys. Rev. {\bf 96}, 99 (1954); T. Kasuya,
Prog. Theor. Phys. {\bf 16}, 45 (1956); K. Yosida, Phys.
Rev. {\bf 106}, 893 (1957).

\bibitem{Taylor} K. H. R.Taylor and M. Darby, Physics of Rare
Solids, Chapman and Hall, London, 1972.

\bibitem{Maki} K. Maki, Phys. Rev. B {\bf 8}, 191 (1973).

\bibitem{Anderson} P. W. Anderson and H. Suhl, Phys. Rev. {\bf
116}, 898 (1969).

\bibitem{Maple} M. B. Maple, Appl. Phys. {\bf 9}, 179 (1976).

\bibitem{Roth} S. Roth, Appl. Phys. {\bf 15}, 1 (1978).

\bibitem{Gorkov} L. P. Gor'kov, Zh. Eksp. Teor. Fiz. {\bf 37},
1407 (1959) [ Sov. Phys. JETP {\bf 10}, 998 (1960)].

\bibitem{Kochelaev} B. I. Kochelaev, L. R. Tagirov, and M. G.
Khusainov Zh. Eksp. Teor. Fiz. {\bf 76}, 578 (1979) [Sov.
Phys. JETP {\bf 49}, 291 (1979)].

\bibitem{BedMul} J. C. Bednorz and K. A. M\"{u}ller, Z. Phys.
{\bf 64}, 189 (1986).

\bibitem{Chu} M. K. Wu, J. R. Ashburn, C. J. Torng, P. H. Hor., R.
L. Meng, L. Gao, Z. J. Huang, Y. Z. Wang, and C. W. Chu, Phys.
Rev. Lett. {\bf 58}, 908 (1987).


\bibitem{Cava} R. J. Cava, B. Batlogg, R. B. van Dover, D.~W.~Murphy, S.~Sunchine, T.~Segrist,
J.~P.~Remeika, E.~A. Rietman, S.~Zahurack, And G.~P.~Espinosa, Phys. Rev. Lett. {\bf
58}, 1676 (1987).

\bibitem{Kaji} C. Kaji, S.~Koriyama, and S.~Nagano, J. Cer. Soc. Jpn. {\bf 96}, 433
(1988).


\bibitem{NE5} N. E. Alekseevskii, I. A. Garifullin, N. N.
Garif'yanov, B. I. Kochelaev, A. V. Mitin, V. I. Nizhankovskii, L.
R. Tagirov, G. G. Khaliullin, and E. P. Khlybov, Pis'ma v Zh.
Eksp. Teor. Fiz. {\bf 48}, 36 (1988) [JETP Lett. {\bf 48}, 37
(1988)].

\bibitem{NE6} N. E. Alekseevskii, A. V. Mitin, V. I. Nizhankovskii, I. A. Garifullin,
N. N. Garif'yanov, G. G. Khaliullin, E. P. Khlybov, B. I. Kochelaev,
L. R. Tagirov, J.Low Temp.Phys. {\bf 77}, 87 (1989).

\bibitem{Mei} Yu. Mei, C. J. Jiang, S. M. Green, H. L. Luo, and C.
Politis, Z.~Phys. {\bf 69}, 11 (1987).

\bibitem{Blombergen} N. Blombergen, J. Appl. Phys. {\bf 23},
1383 (1952).

\bibitem{Tranquada} J. M. Tranquada, A. H. Moudden, A. I. Goldman, P. Zolliker,
P. E. Cox, and G. Shirane, Phys. Rev. B {\bf 38}, 2477 (1988).

\bibitem{Caponi} J. J. Caponi, C. Challout, A. W. Hewat, P. Lejay, M. Marezio,
N. Nguen, B. Raveau, J. L. Soubeyroux, J. L. Tholence, and R.
Tournier, Europhys. Lett. {\bf 3}, 130 (1987).

\bibitem{Kuiper} P. Kuiper, G. Kruzinga, J. Ghijsen, M. Grioni, P. J. W. Weijs,
F. M. F. de Groot, G. A.  Sawatzky, H. Verweij, L. F. Feiner, and H.
Petersen, Phys. Rev. B {\bf 38}, 6483 (1988).

\bibitem{Khachaturyan1} A. G. Khachaturyan, S. V. Semenovskaya, and J. W. Morris, J.r.,
Phys. Rev. B {\bf 37}, 2243 (1988).

\bibitem{Khachaturyan2} A. G. Khachaturyan and J. W. Morris, Jr., Phys. Rev. Lett.
{\bf 61}, 215 (1988).

\bibitem{Orenstein} J. Orenstein and A. J. Mills, Science {\bf 288}, 468 (2000).

\bibitem{Anderson1} P. W. Anderson, Science {\bf 288}, 480 (2000).

\bibitem{Dagotto} E. Dagotto, Science {\bf 309}, 257 (2005).

\bibitem{Bonn} D. A. Bonn, Nature Physics {\bf 2}, 159 (2006).

\bibitem{Elschner} B. Elschner and A. Loidl,  "Electron-spin resonance on localized magnetic moments in metals",
Handbook on the Physics and Chemistry of Rare Earths Vol. 24, 1997, pp. 221-337.

\bibitem{Janossy97} A. J\'{a}nossy, T. Feh\'{e}r, G. Oszl\'{a}nyi, and G. V. M. Williams, Phys. Rev. Lett. {bf 79}, 2726 (1997).

\bibitem{Janossy00} T. Feh\'{e}r, A. J?nossy, G. Oszl\'{a}nyi, F. Simon, B. Dabrowski,
P. W. Klamut, M. Horvati\'{c}, and G. V. M. Williams, Phys. Rev. Lett. {\bf  85}, 5627 (2000).

\bibitem{Kataev97} V. Kataev, B. Rameev, B. B\"{u}chner, M. H\"{u}cker and R. Borowski,
Phys. Rev. B{\bf 56}, R3394 (1997).

\bibitem{Kataev98} V. Kataev, B. Rameev, A. Validov, B. B\"{u}chner, M. H\"{u}cker and R. Borowski,
Phys. Rev. B {\bf 58}, R11876 (1998).

\bibitem{Elschner94} B. I. Kochelaev, L. Kan, B. Elschner, and S. Elschner, Phys. Rev. B {\bf 49}, 13106 (1994).

\bibitem{Elschner97} B. I. Kochelaev, J. Sichelschmidt, B. Elschner, W. Lemor, and A. Loidl,
Phys. Rev. Lett. {\bf 79}, 4274 (1997).


\bibitem{Janossy01} F. Simon, A. J\'{a}nossy, T. Feh\'{e}r, F. Mur\'{a}nyi, S. Garaj, L. Forr\'{o},
C. Petrovic, S. L. Bud'ko, G. Lapertot, V. G. Kogan, and P. C. Canfield, Phys. Rev. Lett. {\bf 87}, 047002 (2001).

\bibitem{Kamihara08} Y. Kamihara, T. Watanabe, M. Hirano, and H. Hosono,  J. Am. Chem. Soc. {\bf 130}, 3296 (2008).

\bibitem{Pascher10} N. Pascher, J. Deisenhofer, H.-A. Krug von Nidda, M. Hemmida, H. S. Jeevan,
P. Gegenwart, and A. Loidl, Phys. Rev. B {\bf 82}, 054525 (2010).

\bibitem{Dengler12} H.-A. Krug von Nidda, S. Kraus, S. Schaile, E. Dengler, N. Pascher, M. Hemmida,
M. J. Eom, J. S. Kim, H. S. Jeevan, P. Gegenwart, J. Deisenhofer, and A. Loidl, Phys. Rev. B {\bf 86}, 094411 (2012).

\bibitem{Alfonsov11} A. Alfonsov, F. Muranyi, V. Kataev, G. Lang, N. Leps, L. Wang, R. Klingeler, A. Kondrat,
C. Hess, S. Wurmehl, A. Koehler, G. Behr, S. Hampel, M. Deutschmann, S. Katrych, N.D. Zhigadlo, Z. Bukowski, J. Karpinski,
B. B\"{u}chner, Phys. Rev. B {\bf 83}, 94526 (2011).

\bibitem{Alfonsov12} A. Alfonsov, F. Muranyi, N. Leps, R. Klingeler, A. Kondrat, C. Hess, S. Wurmehl, A. Koefiler,
G. Behr, V. Kataev, B. B\"{u}chner, JETP {\bf 114}, 662 (2012).

\bibitem{BB} A. I. Buzdin, L. N. Bulaevskii, Adv. Phys. {\bf 34}, 175 (1985).


\bibitem{Garifullin} I. A. Garifullin, D. A. Tikhonov, N. N. Garif'yanov, M. Z.
Fattakhov, K. Theis-Br\"{o}hl, K. Westerholt, and H. Zabel, Appl.
Magn. Reson. {\bf 22}, 439 (2002).

\bibitem{Salikhov1} R. I. Salikhov, I. A. Garifullin, N. N. Garif'yanov,
L. R. Tagirov, K. Theis-Br\"{o}hl, K. Westerholt, and H. Zabel, Phys. Rev. Lett.
{\bf 102}, 087003 (2009).

\bibitem{Salikhov2} R. I. Salikhov, N. N. Garif'yanov, I. A. Garifullin
L. R. Tagirov, K. Westerholt and H. Zabel, Phys. Rev. B {\bf
80}, 214523 (2009).

\bibitem{note}Such calculations have been also performed by Orbach
\cite{Orbach3}.

\bibitem{Orbach3} R. Orbach, Phys.~Lett.~A {\bf 47}, 281 (1974).


\end{thebibliography}
\end{document}